\documentclass[twoside,english,twocolumn,5p,authoryear]{elsarticle}
\usepackage[T1]{fontenc}
\usepackage{color}
\usepackage{babel}
\usepackage{graphicx}
\usepackage[unicode=true,
 bookmarks=true,bookmarksnumbered=false,bookmarksopen=false,
 breaklinks=false,pdfborder={0 0 0},backref=false,colorlinks=true]
 {hyperref}

\makeatletter

\providecommand{\tabularnewline}{\\}

\journal{Astronomy and Computing}

\usepackage{subfig}
\usepackage{hyperref}
\usepackage{natbib}
\usepackage{xspace}

\makeatother

\begin{document}

\begin{frontmatter}{}

\title{HDFITS: porting the FITS data model to HDF5 }

\author[cfa]{D. C.~Price\corref{cor1}}

\ead{dprice@cfa.harvard.edu}

\author[cfa]{B. R.~Barsdell}

\author[cfa]{L. J.~Greenhill}

\ead{greenhill@cfa.harvard.edu}

\author{}

\cortext[cor1]{Corresponding author}

\address[cfa]{Harvard-Smithsonian Center for Astrophysics, MS 42, 60 Garden Street,
Cambridge MA 01238 USA}
\begin{abstract}
The FITS (Flexible Image Transport System) data format has been the
de facto data format for astronomy-related data products since its
inception in the late 1970s. While the FITS file format is widely
supported, it lacks many of the features of more modern data serialization,
such as the Hierarchical Data Format (HDF5). The HDF5 file format
offers considerable advantages over FITS, such as improved I/O speed
and compression, but has yet to gain widespread adoption within astronomy.
One of the major holdbacks is that HDF5 is not well supported by data
reduction software packages and image viewers. Here, we present a
comparison of FITS and HDF5 as a format for storage of astronomy datasets.
We show that the underlying data model of FITS can be ported to HDF5
in a straightforward manner, and that by doing so the advantages of
the HDF5 file format can be leveraged immediately. In addition, we
present a software tool, \texttt{\normalsize{}fits2hdf}, for converting
between FITS and a new `HDFITS' format, where data are stored in HDF5
in a FITS-like manner. We show that HDFITS allows faster reading of
data (up to 100x of FITS in some use cases), and improved compression
(higher compression ratios and higher throughput). Finally, we show
that by only changing the import lines in Python-based FITS utilities,
HDFITS formatted data can be presented transparently as an in-memory
FITS equivalent. \end{abstract}

\end{frontmatter}{}

\section{Introduction}

The Flexible Image Transport System (FITS) file format has enjoyed
several decades of widespread usage within astronomy \citep{1979ipia.coll..445W,1980SPIE..264..298G}.
The ubiquity enjoyed by FITS has been attributed in part to the guiding
maxim ``once FITS, always FITS'': that changes to the FITS standard
must be incremental so as to never break backward compatibility \citep{2003ASSL..285...71G}.
For this reason -- among others -- it is familiar to many generations
of astronomers, and the large ecosystem of software that has been
created over the years has in turn motivated further adoption of the
standard. In particular, the \texttt{CFITSIO} library (\href{http://ascl.net/1010.001}{ascl:1010.001},
\citealp{1999ASPC..172..487P,2010ascl.soft10001P}) for the reading
and writing of FITS files has become the \emph{de facto} standard.

FITS has necessarily evolved over the years, with the addition of
features such as random groups \citep{1981A&AS...44..371G}, ASCII
tables \citep{Harten1998}, binary tables \citep{Cotton1995}, and
compression \citep{2002SPIE.4847..444P,2007ASPC..376..483S,2009PASP..121..414P,2010ascl.soft10002S}.
By culmination of these additions, the FITS file format is now officially
at version 3.0 \citep{2010A&A...524A..42P}. However, these changes
have been relatively minor iterations upon the core FITS format. The
``once FITS, always FITS'' maxim limits what modifications can be
made; the guiding principle that has made FITS so successful can now
be seen as limiting its applicability.

The limitations of FITS are succinctly summarized in \citet{2014ASPC..485..351T}
and \citet{Thomas:2014vq}. As the size of data products increase,
new paradigms for data processing become increasingly important \citep{Kitaeff:2014ua}.
For example, the Large-Aperture Experiment to Detect the Dark Ages
(LEDA, \citealp{Greenhill:2012um}) produces 24TB per day, and the
Canadian Hydrogen Intensity Mapping Experiment (CHIME, \citealp{Bandura2014}),
produces 4TB /day in its pathfinder alone. Future telescopes, such
as the Square Kilometre Array%
\footnote{\url{http://www.skatelescope.org}%
} (SKA) will produce over 10x the current global internet traffic \citep{ska2015}.
Distributing such massive data volumes is prohibitive and requires
significant amounts of data reduction to be done in real-time, with
high-throughput data compression and massively parallel data access.
FITS is not well-equipped to deal with these challenges.

Several authors have proposed alternative serializations that have
advantages over FITS. In \citet{Kitaeff:2014ua}, the authors consider
JPEG2000 as an alternative format for images and data cubes. \citet{thomas2001}
discuss advantages of converting FITS files to XML; \citet{Jennings1995}
considered HDF4 (Hierarchical Data Format) as a format . Work toward
an HDF5-based format for astronomy was proposed by \citet{2011ASPC..442..663W},
but funding for this was not secured. 

Motivated by data volumes, HDF5 has also been proposed or has been
implemented for for the LOFAR radio telescope \citep{Anderson:2010tp},
the CCAT telescope \citep{Schaaf:2014}, the CHIME pathfinder \citep{masui2015},
and MeerKat telescope \citep{meerkat2015}, among others. These implementations
share a common file format, but the data is organized in differing
ways as there is no agreed-upon method.

Here, we discuss the immediate, practicable advantages of HDF5 as
an alternative serialization format for the FITS data model. We show
that data model inherent to the FITS file format can be converted
in a straightforward manner to HDF5, and that by doing so better compression
results and faster read speeds can be achieved. This work extends
that presented in proceedings of Astronomical Data Analysis Software
and Systems (ADASS) XXIV \citep{Price:2014ue}.

\subsection{Definitions}

In order to discuss data storage methods and formats without ambiguity,
we first need to clarify our vocabulary: 
\begin{itemize}
\item \emph{Data model}: a high-level, conceptual model of data, types of
data, and how data are organized, e.g. `group' and `dataset'. 
\item \emph{Data schema}: a lower-level, domain-specific ontology (i.e.
framework that gives meaning) of how data and metadata are arranged
inside a data model. For example, a schema may define a set of rules
for the names of attributes and datasets, and how data are organized
within the data model. 
\item \emph{Serialization, or storage model}: how objects from the data
model are mapped to bytes within an address space on storage media
(such as a hard drive). 
\item \emph{File format}: a well-defined serialization for a data model.
\item \emph{Convention}: a documented data schema that has widespread acceptance
within a community of users. 
\item \emph{Standard}: the acknowledged, formal specification of a file
format. A standard may or may not define acceptable data models and
schema, but should provide an application programming interface API.
\end{itemize}
From this view, the data model can be seen as\emph{ syntax}, while
the data schema may be seen as the ontology that gives \emph{semantics}.
Without a well-defined schema, the underlying meaning of the dataset
may be unclear. Neither the FITS nor HDF5 standards define data schema;
however there are registered FITS conventions%
\footnote{\url{http://fits.gsfc.nasa.gov/fits_registry.html}%
} for certain classes of data. For FITS files, the data model is closely
tied to the storage model; in contrast HDF5 allows abstract data models
that are divorced from the subtleties of the storage model.

\subsection{Motivation}

The goal of the work presented here is to create an HDF5-based equivalent
of the FITS file format, and to provide utilities for converting between
the two formats. The motivation of this approach, as opposed to creating
an HDF5-based format from scratch, is that decades of widespread FITS
usage has left a legacy that would otherwise be discarded. By preserving
the familiar underlying data model of FITS, software packages designed
to read and interpret FITS can be readily updated to read HDF5 data.
Maintaining backwards-compatibility with FITS, so that data stored
in HDF5 files can be converted into FITS for use in legacy software
packages is another persuasive reason to pursue a FITS-like data model
within HDF5.

A switch to an HDF5-based format has several advantages, many of which
are immediately practicable. Perhaps the most compelling for next-generation
datasets is that HDF5 supports far more compression filters than FITS;
some comparisons of compression are shown in Section \ref{sec:Performance-comparison}.
For the data tested here, HDF5 compressors outperform FITS equivalents,
although we note a larger cross-section of astronomical data must
be compared to form definitive conclusions. Another compelling reason
to switch to HDF5 is I/O speed. HDF5 allows for efficient reading
of portions of a dataset, whether they are contiguous or a regular
pattern of points or blocks. Additionally, HDF5 has parallel I/O support,
which is becoming increasingly important for efficient processing
of large datasets on multi-node systems. 

A good example of porting a data model is provided by \citet{Jenness:2015bi},
in which the Hierarchical Data System (HDS) format has been reimplemented
in HDF5; similarly, \citet{Jenness:2014wy} discusses conversion to
and from FITS and the HDS-based NDF (N-dimensional data format). Together,
the ability to convert FITS to NDF and the reimplementation of HDS
within HDF5, provides an alternative path toward conversion of FITS
to a HDF5-based format. A comparison of the two approaches is given
in Section \ref{sec:Discussion}.

A port of the FITS data model to HDF5 does not, however, address issues
with the FITS data model itself.  Nevertheless, as the HDF5 data
model is abstracted from its file format, an HDF5-based version of
the FITS data model can be extended without requiring changes to the
storage model. The HDF5-based FITS equivalent, as detailed here, can
be used as a starting point and as a testbed for enhancing the FITS
data model. We provide a utility using the \texttt{\textsl{\emph{fits2hdf}}}
utility (described in section \ref{sec:fits2hdf-software}), a user
can convert their data into HDF5 and convert it back into FITS if
required. Our hope is that this provides a means for the astronomy
community to investigate the advantages and disadvantages of moving
away from FITS.

\subsection{Overview}

This article is organized as follows. In Section \ref{sec:Comparison-of-FITS},
we provide a comparison of the FITS and HDF5 file formats, with emphasis
on their data models. In Section \ref{sec:Porting-FITS} we present
a mapping of the FITS data model into the HDF5 abstract data model
we call `HDFITS'; section \ref{sec:fits2hdf-software} details software
with which to convert FITS to and from the HDFITS format. A comparison
of compression and read speed on equivalent datasets in both formats
is then given in Section \ref{sec:Performance-comparison}. This is
followed by a discussion of the benefits and then some concluding
remarks.

\section{Comparison of FITS and HDF5 data models\label{sec:Comparison-of-FITS}}

One of the main differences between FITS and HDF5 is that FITS does
not abstract the data model from the storage model; that is, there
is a simple correspondence between the data model and the serialization.
A FITS file is composed of a set of Header-Data Units (HDUs), which
are ASCII headers followed by contiguous blocks of data (binary or
ASCII encoded). In comparison, an HDF5 file is organized as a directed
graph, and objects from the data model are mapped to the storage model.
By use of B-tree data structures, data may be discontiguous, allowing
insertions and resizing of datasets. HDF5 also allows `hierarchy',
whereby an object can be placed within a group, and nested groups
of objects are allowed.

Here, we present a brief comparison of the two data models; we refer
the reader to \citet{2010A&A...524A..42P} and \citet{hdfgroup2015}
for further details of the two file formats.

\subsection{FITS data model}

The data model of FITS is closely related to the the FITS file format
itself. Over the years, changes to the format have allowed the data
model to evolve, without significant change to the storage model.
For example, header keywords may now be longer than 8 characters,
but during serialization keywords are stored in a way that each header
in the keyword remains shorter than 8 characters. Here, we give a
short overview of FITS, with an emphasis on the data model.

\paragraph{Header data unit}

A FITS file is comprised of segments known as Header-Data Units (HDUs).
Each HDU consists of an ASCII `header unit' consisting of keyword-value
pairs (known as 'cards'), and may be followed by an optional 'data
unit'. The header unit consists of metadata that describes the structure
of the data unit and the contents of the file. At a minimum, a FITS
file contains one HDU, which is referred to as the `Primary HDU';
HDUs after this are referred to as `Extension HDUs'. A file with multiple
HDUs is referred to as a Multi-Extension FITS (MEF), otherwise it
is known as a Single-Image FITS (SIF). Due to the implementation of
FITS compression, a SIF file must be converted into a MEF file in
order to apply the compression filter. As such, SIF files are becoming
less common.

\paragraph{Header unit}

\begin{figure}
\texttt{\textcolor{blue}{\footnotesize{}SIMPLE}}\texttt{\scriptsize{}
~= ~~~~~T / file conforms to FITS standard }~\\
\texttt{\textcolor{blue}{\scriptsize{}BITPIX}}\texttt{\scriptsize{}
~~= ~~~~16 / number of bits per data pixel }~\\
\texttt{\textcolor{blue}{\scriptsize{}NAXIS}}\texttt{\scriptsize{}
~~~= ~~~~~2 / number of data axes }~\\
\texttt{\textcolor{blue}{\scriptsize{}NAXIS1}}\texttt{\scriptsize{}
~~= ~~~440 / length of data axis 1 }~\\
\texttt{\textcolor{blue}{\scriptsize{}NAXIS2}}\texttt{\scriptsize{}
~~= ~~~300 / length of data axis 2}{\scriptsize \par}

\protect\caption{Example of a FITS header unit, showing keyword, value, and comment
structure. Here, whitespace has been trimmed.\label{fig:header-example}}

\end{figure}

The FITS header unit is an ASCII-formatted list of keyword-value pairs
and short associated comments (Fig\texttt{\footnotesize{}~\ref{fig:header-example}}).
These keyword-value pairs are used to describe and document the data
contained within the data unit (if present); that is, they are metadata.
For example, a header unit may contain labels to array dimensions,
coordinate system information, and information about the instrument
from which the data originate. Depending upon the type of data unit,
there are some mandatory keywords that must be present.

Within a FITS file, each line in the header must be no more than 80
characters long, with each keyword in all-caps and under 8 characters
long; corresponding values must be no longer than 68 characters. This
can be considered a serialization quirk, as the data model has been
extended by the use of special keywords, with two common variants:
the {\tt HIERARCH} keyword can be used to allow keywords up to 64 characters
long, and the {\tt CONTINUE} keyword allows values to span over multiple
lines. Each entry in the header unit is referred to as a 'card', so
that the header unit can be considered an ordered list of cards.

Two other cards that may be present within the header unit are the
{\tt COMMENT} and {\tt HISTORY} cards, which allow plaintext comments and notes
about the file's history. Long comments and history are created via
multiple cards; again, this is more a serialization detail than an
aspect of the data model.

\paragraph{Data Unit}

There are three classes of data unit (known as `extensions') that
may be stored in FITS: the {\tt IMAGE} extension, which stores images
and N-dimensional data; {\tt TABLE}, which is used to store ASCII-formatted
tables; and {\tt BINTABLE}, which stores tables in a more efficient binary
format and unlike the {\tt TABLE} extension can store arrays of data. There
is also a `random group' data unit which is now deprecated but still
used for radio interferometer data, for historical reasons. If compression
is applied to an {\tt IMAGE} extension, it is converted into a {\tt BINTABLE}.

\paragraph{Datatypes}

The type of data within an HDU is specified in mandatory header cards,
and is limited to: 8-bit unsigned integers; 16, 32 and 64-bit signed
integers; 32 and 64-bit IEEE 754 floating point; and 7-bit ASCII (ANSI
1977) data. Boolean and bit data may also be stored. With the exception
of 8-bit data serialization of unsigned integers is not supported,
but this may be circumvented by the use of an accompanying scale offset
keyword ({\tt BZERO}).

\paragraph{World Coordinate Systems}

An integral part of FITS is its ability to store metadata detailing
the mapping between coordinates within an image and physical (i.e.,
world) coordinate systems (WCS). Coordinate systems are specified
via keywords in the header unit and must follow the coordinate system
definitions \citep{Calabretta:2002gz,Greisen:2002id,Calabretta:2004wr,Greisen:2005jp,Rots:2014iw}.
WCS information allows FITS viewers to interpret the coordinates that
correspond to each pixel and thus overlay graticules and other information.

\paragraph{FITS conventions}

Higher-level data models do exist for FITS in the form of 'conventions'%
\footnote{A registry of FITS conventions is available at \url{http://fits.gsfc.nasa.gov/fits_registry.html}%
}, which prescribe a set of header keywords, values, and table/image
structures, generally for a domain-specific application. For example,
UVFITS \citep{Greisen:2012tw} and FITS-IDI \citep{Griesen:2008tn}
are conventions for the storage of data from radio interferometers,
and the SDFITS convention \citep{garwood2000} was designed for storage of data from
single dish observations.

\subsection{HDF5 data model}

HDF5 employs an abstract data model that is designed to conceptually
cover many different models. Unlike FITS, HDF5 allows hierarchy within
files, meaning for example that a group may be contained within a
group. Here, we briefly introduce aspects of HDF5 that are directly
relevant when considering how to port the FITS data model into HDF5;
the HDF5 data model is described in further detail in \citet{hdfgroup2015}.

\paragraph{Group}

An HDF5 group is simply a collection of objects (a group is itself
an object). Each HDF5 file contains a root group, which may contain
zero or more other groups. Every object within an HDF5 file -- apart
from the root group -- must be a member of at least one group. Conceptually,
a FITS HDU can be considered a group containing a header unit and
a data unit.

\paragraph{Dataset}

An HDF5 dataset is a multidimensional array of elements of a given
data type. A dataset could consist of a simple datatype (e.g. integers),
or a composite datatype consisting of several different kinds of data
element. an HDF5 dataset is similar to a FITS data unit.

\paragraph{Datatype}

The HDF5 datatype is a description of a specific class of data element.
Atomic datatype elements include strings, floats, integers and bitfields.
Composite datatypes, such as an array, are formed by combining multiple
atomic data elements. HDF5 also allows for users to define custom
datatypes, for example a 17-bit integer. All the datatypes supported
by FITS are included within the standard predefined HDF5 datatypes.

\paragraph{Attributes}

HDF5 attributes are similar to FITS header cards, in that they are
a keyword - value metadata pair used to describe data. Both HDF5 groups
and datasets may have attributes attached to them. HDF5 attributes
differ to FITS header cards in that the data stored in an attribute
may be composite datatypes such as arrays; further, HDF5 attributes
are not stored as an ordered list.

\paragraph{Dataspace}

An HDF5 dataspace is a description of the dimensions of a multidimensional
array. A dataspace provides a similar description of array dimensions
to the cards within a FITS header that detail dimensions.

\paragraph{Dimension scales}

An HDF5 dimension scale is a 1-dimensional dataset that provides information
about the dimensions of a given dataspace. This is analogous to the
WCS-related cards within a FITS header.

\paragraph{Existing specifications}

There are two HDF5 specifications that are relevant for the storage
of astronomy data: the {\tt IMAGE} \citep{hdfimage2015} and {\tt TABLE} \citep{hdftable2015}
specifications. These documents provide a standard method for storing
image and tabular data, respectively. These specifications define
mandatory attributes to define their properties. They are distinguished
by other datasets via the attribute {\tt CLASS}.

FITS has shown that it is possible to store almost all astronomy datasets
in either a table or as an N-dimensional image. While this may be
true, there is likely significant advantage to defining further specifications
(new classes or even sub-classes), that may be more appropriate and
further provide more semantic meaning. We suggest that classes and
subclasses should remain abstract, such as `time series' or `sparse
matrix', in contrast to higher-level conventions for domain-specific
data (e.g. `single dish observation').

\section{Porting FITS to HDF5\label{sec:Porting-FITS}}

\begin{figure*}[t]
\begin{centering}
\includegraphics[width=1.6\columnwidth]{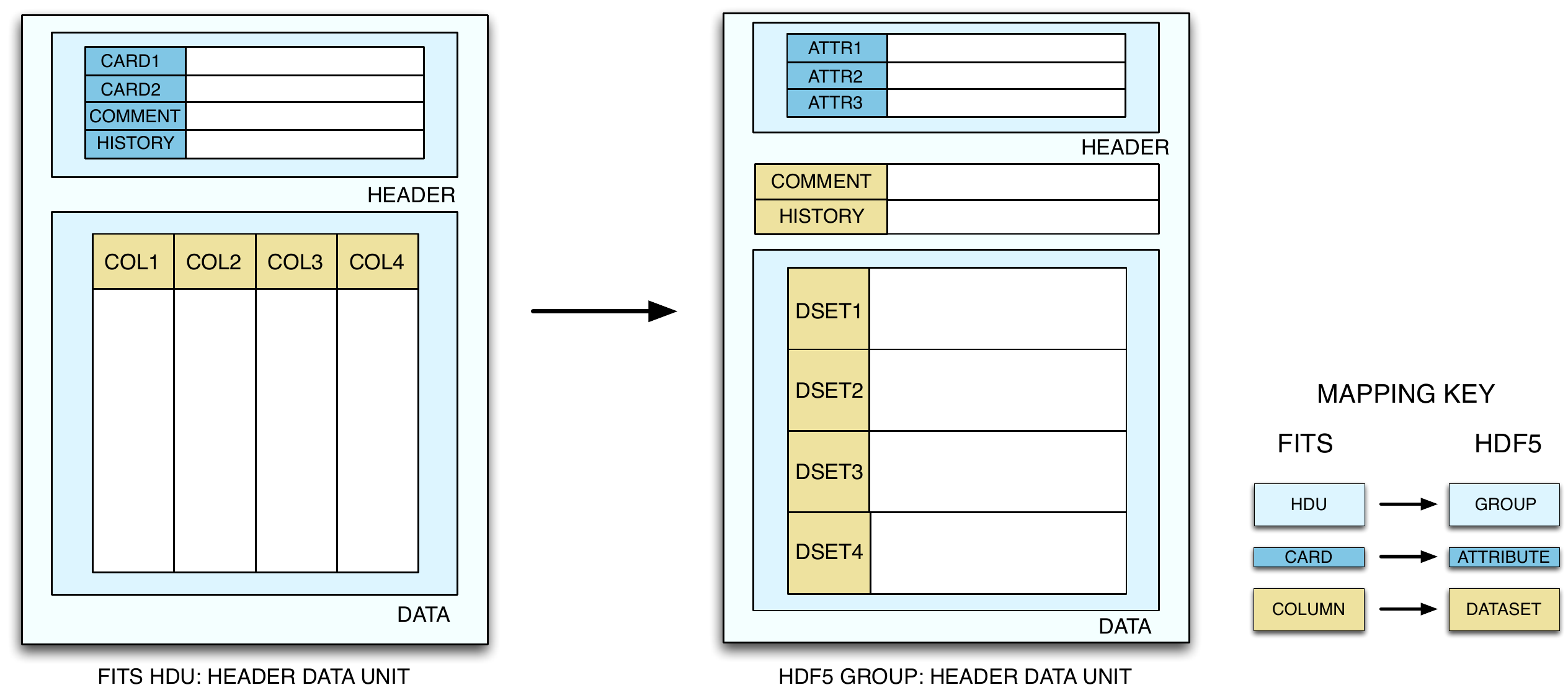}
\par\end{centering}

\protect\caption{Diagram showing the mapping of the FITS data structure (for a binary
table) into the HDF5 data model, where each column is stored as a
dataset within a group. Reproduced with modifications from \citep{Price:2014ue}.
\label{fig:fits-mapping}}
\end{figure*}
 There are myriad ways in which FITS data could be stored within
HDF5. We will use the portmanteau `HDFITS' to refer to data stored
in HDF5 with a FITS-like data model. In order to port the FITS data
model to HDF5, we first need to decide upon how best to create an
HDU-like object within HDF5. There are a number of possible approaches,
such as:
\begin{itemize}
\item \emph{Single `HDU' dataset with attributes. }A HDU is created from
a single dataset object. The values contained in the FITS header unit
are mapped to attributes attached to the dataset, and the data payload
is stored in the dataset's dataspace. Comments and history are also
stored as attributes.
\item \emph{A `HDU' group with a `header' dataset and a `data' dataset.}
A HDU-like object is created by placing two datasets within a group.
Header values, comments, and history are stored within a table in
the 'header' dataset's dataspace, and the main data payload is stored
in the dataset's dataspace.
\item \emph{A `HDU' group with a `header' attributes and a `data' group.}
A HDU-like object is created by placing a child dataset (or group)
within a parent group. Header values are stored in the parent group's
attributes; 'comment' and 'history' datasets are also placed within
the parent group. A number of datasets may be placed within the 'data'
group; for example, each column in a table could be stored as a dataset.
\end{itemize}
We have implemented the latter (Figure \ref{fig:fits-mapping}), as
we have found it to be intuitive, while allowing flexibility for the
data model to be extended in the future by addition of other datasets
or groups.

The next step is to define how HDU objects are arranged within the
HDF5 file. As HDF5 allows hierarchy, it would allow groups of related
HDUs to be stored; we elected to enforce a that all HDUs are located
in the root group, as otherwise the data model would not be compatible
with FITS. Unlike FITS, HDF5 does not enforce ordering of groups%
\footnote{The default ordering is increasing lexicographic by attribute name.
One might instead track by attribute / link creation time, but this
is less transparent.%
}; to reproduce this each group must have an attribute identifying
its position within the HDU list. 

\begin{figure}
\begin{centering}
\includegraphics[width=0.99\columnwidth]{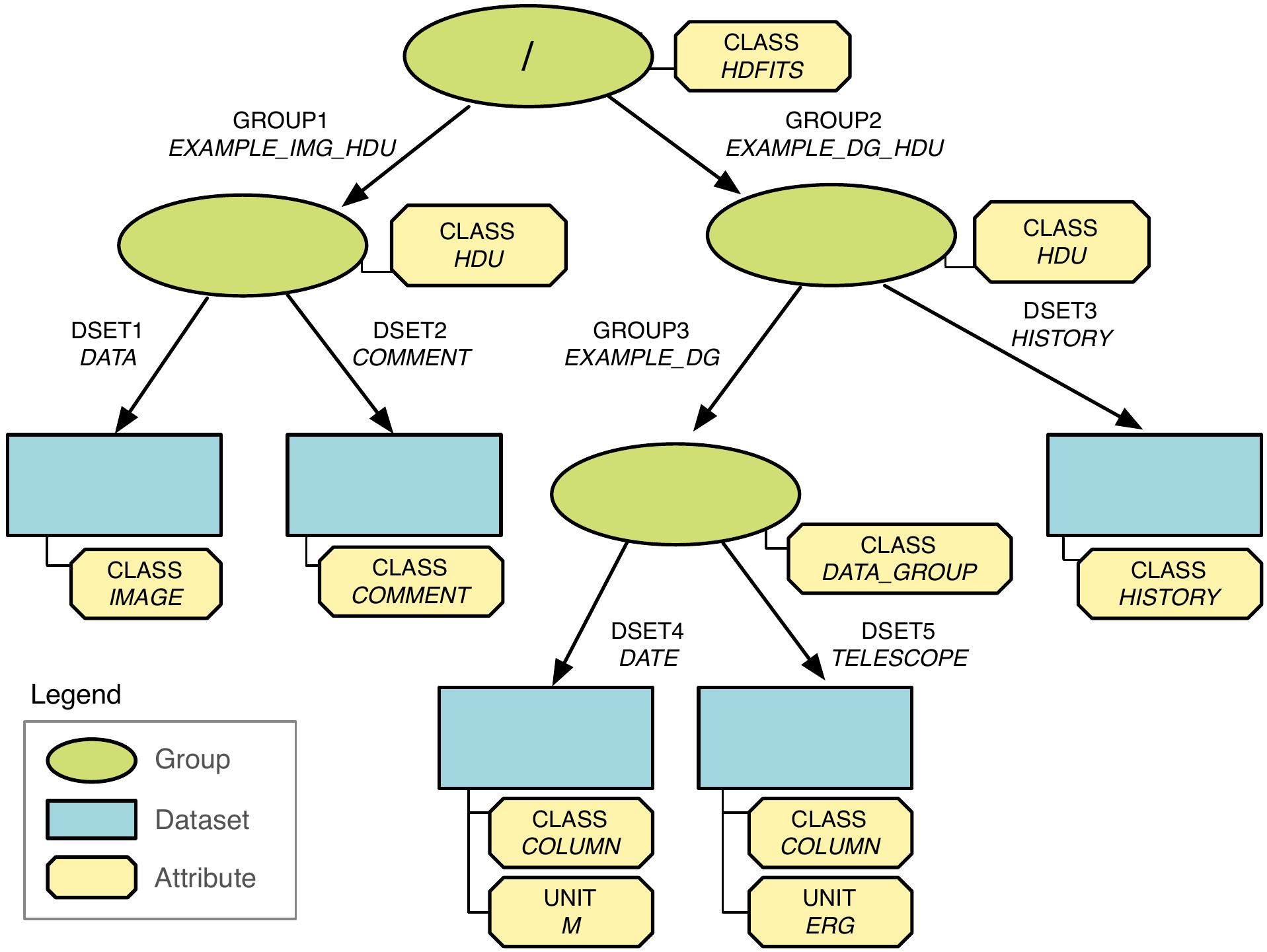}
\par\end{centering}

\protect\caption{A tree diagram showing an example implementation of a HDFITS data
model. \label{fig:tree-diagram}}
\end{figure}

We use the attribute {\tt CLASS=HDU} to identify header data units, and
the attribute {\tt CLASS=HDFITS} in the root of the file to identify the
file as a HDFITS file. We further use the keyword CLASS to distinguish
images and table datasets. There is no need to constrain attribute
keywords to be all-caps and 8 characters in HDF5, but for backwards
compatibility we suggest that this is appropriate, at least for a
first HDFITS revision. The structure of an example HDFITS file is
shown in Figure$~$\ref{fig:tree-diagram}. Note that in HDF5, objects
may have multiple names, so groups and datasets are assigned names
by the links that reference them; we enforce that each group and dataset
has a single name that identifies it.

HDF5 dimension scales provide a more flexible way of describing coordinate
systems than the FITS WCS header keywords; this is as entire datasets,
complete with their own attributes, may be linked to label the scale
of each dimension. Since these are 1 dimensional, this is still insufficient
for cases where pixel mappings are not parallel to the axes (for example,
spherical coordinates). In such a case, a parallel array of dimensional
mappings would be more appropriate. For backwards-compatibility with
FITS, we have not implemented dimension scales or further deviations
in this initial version of HDFITS.

\subsection{FITS headers}

There are a number of approaches one can take to implementing an equivalent
of the FITS header within HDF5; we opted to map them to HDF5 attributes.
One could also consider storing the header in a dataset, perhaps even
maintaining the 80-character per line card structure. The latter approach
would be advantageous for `round-tripping' --- conversion from FITS
to HDF5 and back again --- as the header could be kept identical and
intact. The disadvantage is precisely that this does not parse the
header into HDF5 attributes, which means knowledge of how to parse
the FITS header is required for understanding of the corresponding
data unit. Creation of HDFITS files from scratch would also require
creation and storage of superfluous FITS header cards.

Another approach would be to store keyword, value, and comment as
columns of a tabular dataset. Nonetheless, this still requires understanding
and parsing the FITS header. It remains that it is more in keeping
with the HDF5 abstract data model to store metadata in attributes,
hence our implementation. Further discussion of round-tripping is
given in Section \ref{sub:Round-trip-FITS-conversion} below.

\subsection{Image HDUs}

There is an existing HDF5 {\tt IMAGE} specification \citep{hdfimage2015}
for the storage of image data, which we reuse here for the storage
of n-dimensional datasets as are stored in FITS {\tt IMAGE} HDUs. An image
dataset is distinguished by an attribute {\tt CLASS=IMAGE} attached to
the dataset. We note that semantically, much data stored in FITS image
HDUs is not an image at all, but rather an N-dimensional dataset.
Future implementations of HDFITS may elect to create a class {\tt NDDATA},
and to differentiate between different kinds of data (e.g. images,
spectral cubes), using a {\tt SUBCLASS} attribute.

\subsection{Table HDUs}

An HDF5 specification for the storage of data tables already exists
\citep{hdftable2015}, but we have chosen an alternative implementation,
where {\tt CLASS=COLUMN} datasets are stored with a group with attribute
{\tt CLASS=DATA\_GROUP}. Each column within a {\tt DATA\_GROUP} has attributes
to describe its data, such as its position in the table ({\tt POSITION}
attribute), and the units ({\tt UNIT} attribute) of the contained data.
Our motivation for this is that compression algorithms work better
on non-compound datatypes, and also that this allows for columns to
be added and deleted with requiring dataset resizing. Additionally,
data analysis may be orders of magnitude faster if performed on data
held in `column stores' than it can on data held in `row stores' \citep{Abadi:2008gb}. 

The \texttt{fits2hdf} and \texttt{hdf2fits} programs (detailed below),
parse both the HDF5 specified {\tt CLASS=TABLE} and the {\tt CLASS=DATA\_GROUP}
to create an in-memory table object, which can then be written back
to either a {\tt CLASS=TABLE} or {\tt CLASS=DATA\_GROUP} object.

\section{The \emph{fits2hdf} software package\label{sec:fits2hdf-software}}

We have implemented a software package called \texttt{fits2hdf}%
\footnote{\url{https://github.com/telegraphic/fits2hdf}%
},\emph{ }which converts a FITS file into HDFITS files. This utility
is written in Python, and uses the \texttt{astropy} (\href{http://ascl.net/1304.002}{ascl:1304.002},
\citealp{AstropyAA}) library for FITS I/O, and \texttt{h5py}%
\footnote{\url{http://www.h5py.org/}%
} library for HDF5 I/O. As the HDFITS data model is a restricted subset
of the complete HDF5 data model, any HDFITS file may be converted
back into a FITS file without complication; we provide a utility \texttt{hdf2fits}
to provide this functionality.

Internally, \texttt{fits2hdf} converts both HDFITS and FITS data into
an intermediary in-memory data objects. These objects are subclasses
of the \texttt{astropy} \texttt{NDData} and \texttt{Table} classes.
The \texttt{astropy} library is designed to create data objects such
as these that are abstracted from the details of the underlying serialization.
For legacy reasons, the \texttt{astropy} FITS handling routines do
not directly read into these abstract classes, but future releases
intend to rectify this \citep{astropytables2015}. \texttt{Astropy}
does not provide an abstracted `HDU list' object, nor objects to store
comments and history; these are instead defined in \texttt{fits2hdf}.
Future development of \texttt{fits2hdf} will attempt to align with
the development path of \texttt{astropy}, and if possible, all of
the \texttt{fits2hdf} functionality will be transferred to the main
astropy package.

\subsection{Round-trip FITS conversion\label{sub:Round-trip-FITS-conversion}}

A perfectly lossless conversion from FITS into HDFITS, and back into
FITS, should produce an output file that is byte identical to the
input. We stress that this is not the case with \texttt{fits2hdf},
so caution that it should not be used for data archiving without careful
comparison of input and output data. 

There are several reasons as to why input and output files differ.
The first is simply that \texttt{fits2hdf} adds its own comments to
the {\tt HISTORY} table. A second reason is that many FITS files contain
comments on mandatory header cards, such as
\begin{quote}
\texttt{\footnotesize{}SIMPLE = $~$$~$$~$T / This file is a valid
FITS file}{\footnotesize \par}

\texttt{\footnotesize{}BITPIX = $~~$16 / Number of bits per pixel}{\footnotesize \par}
\end{quote}
that encode information meant for the FITS reader, not the user. These
cards, while necessary, are automatically generated when creating
FITS HDUs with \texttt{astropy}; HDF5 does not have an equivalent.
As such, preserving such information is not constructive, and would
require adding useless attributes to the HDF5 file. Put another way,
FITS cards that describe \emph{dataspaces} and \emph{datatypes} are
discarded, and only FITS header cards that provide valuable metadata
to the end user are kept. 

Similarly, table keywords such as {\tt TUNIT} are stripped and
parsed by \texttt{fits2hdf} and their information added to the astropy
column object. Units that are not formatted as suggested in the FITS
standard (e.g. {\tt DEGREES} in lieu of {\tt DEG}), are fixed where possible.

Secondly, some FITS files can store tabular data in either random
groups, which are deprecated in the FITS v3.0 standard, ASCII tables,
or binary tables. The intermediary data format used within \texttt{fits2hdf}
does not differentiate between these different table serializations,
and in export all tabular data is written to the more efficient binary
table extension.

Nonetheless, the values of the data contained within a HDU's data
portion will remain unchanged. We provide a utility called \texttt{fits2fits}
in the root directory of the \texttt{fits2hdf} package, to facilitate
a comparison between a `round-tripped' FITS file.

\section{Performance comparison\label{sec:Performance-comparison}}

Performance comparisons of FITS and HDFITS were conducted on a Macbook
Pro late 2013 model with a solid-state disk. Results in Section \ref{sub:Data-access}
are from data stored on a Western Digital 1TB hard disk (WDC WD1000CHTZ-0).
Both HDF5 and FITS files were read into memory by Python 2.7.6 scripts
which were run within the \texttt{iPython} v1.1 environment. HDF5
files were read with \texttt{h5py} v2.3.1, with \texttt{HDF5} v1.8.13,
and FITS files were read using the \texttt{astropy} v1.0 FITS I/O
library. To test FITS compression, we used the \texttt{FPACK} utility
(\href{http://ascl.net/1010.002}{ascl:1010.002}, \citealp{2010ascl.soft10002S}),
provided as part of the \texttt{CFITSIO} v3.3 package.

As FITS does not support parallel I/O, we use the serial version of
HDF5 for comparison purposes; as such, no parallel read/write tests
were performed. Fair comparison of differing lossy compression schemes
is more involved than we wish to undertake here. As FITS applies lossy
compression to floating point data, we only consider integer data
compression ratios.

\subsection{Compression}

\begin{table*}
\begin{centering}
\begin{tabular}{cccccccccc}
\hline 
\multicolumn{2}{l}{{\footnotesize{}Image details}} & \multicolumn{2}{c}{{\footnotesize{}FITS RICE}} & \multicolumn{2}{c}{{\footnotesize{}HDF5 LZF\_SS}} & \multicolumn{2}{c}{{\footnotesize{}HDF5 Bitshuffle}} & \multicolumn{2}{c}{{\footnotesize{}GZIP}}\tabularnewline
{\footnotesize{}data range ($2^{N}$)} & {\footnotesize{}image size } & {\footnotesize{}ratio } & {\footnotesize{}time (s)} & {\footnotesize{}ratio} & {\footnotesize{}time (s)} & {\footnotesize{}ratio} & {\footnotesize{}time (s)} & {\footnotesize{}ratio} & {\footnotesize{}time (s)}\tabularnewline
\hline 
\hline 
{\footnotesize{}7} & {\footnotesize{}2048 x 2048} & {\footnotesize{}3.52} & {\footnotesize{}0.15} & {\footnotesize{}3.55} & {\footnotesize{}0.12} & \textbf{\footnotesize{}3.84} & \textbf{\footnotesize{}0.04} & {\footnotesize{}2.47} & {\footnotesize{}3.64}\tabularnewline
{\footnotesize{}15} & {\footnotesize{}2048 x 2048} & {\footnotesize{}1.87} & {\footnotesize{}0.17} & {\footnotesize{}1.92} & {\footnotesize{}0.17} & \textbf{\footnotesize{}1.98} & \textbf{\footnotesize{}0.04} & {\footnotesize{}1.46} & {\footnotesize{}3.01}\tabularnewline
{\footnotesize{}23} & {\footnotesize{}2048 x 2048} & {\footnotesize{}1.28} & {\footnotesize{}0.21} & {\footnotesize{}1.33} & {\footnotesize{}0.17} & \textbf{\footnotesize{}1.32} & \textbf{\footnotesize{}0.04} & {\footnotesize{}1.13} & {\footnotesize{}0.82}\tabularnewline
{\footnotesize{}31} & {\footnotesize{}2048 x 2048} & {\footnotesize{}0.99} & {\footnotesize{}0.24} & {\footnotesize{}1.00} & {\footnotesize{}0.12} & \textbf{\footnotesize{}1.00} & \textbf{\footnotesize{}0.05} & {\footnotesize{}1.00} & {\footnotesize{}0.55}\tabularnewline
\hline 
{\footnotesize{}7} & {\footnotesize{}8192 x 8192} & {\footnotesize{}3.54} & {\footnotesize{}2.03} & {\footnotesize{}3.55} & {\footnotesize{}1.82} & \textbf{\footnotesize{}3.84} & \textbf{\footnotesize{}0.49} & {\footnotesize{}2.47} & {\footnotesize{}59.08}\tabularnewline
{\footnotesize{}15} & {\footnotesize{}8192 x 8192} & {\footnotesize{}1.88} & {\footnotesize{}2.55} & {\footnotesize{}1.92} & {\footnotesize{}2.49} & \textbf{\footnotesize{}1.98} & \textbf{\footnotesize{}0.55} & {\footnotesize{}1.46} & {\footnotesize{}49.63}\tabularnewline
{\footnotesize{}23} & {\footnotesize{}8192 x 8192} & {\footnotesize{}1.28} & {\footnotesize{}3.08} & {\footnotesize{}1.33} & {\footnotesize{}2.85} & \textbf{\footnotesize{}1.32} & \textbf{\footnotesize{}0.64} & {\footnotesize{}1.13} & {\footnotesize{}13.91}\tabularnewline
{\footnotesize{}31} & {\footnotesize{}8192 x 8192} & {\footnotesize{}0.99} & {\footnotesize{}3.38} & {\footnotesize{}1.0} & {\footnotesize{}1.95} & \textbf{\footnotesize{}1.00} & \textbf{\footnotesize{}0.68} & {\footnotesize{}1.0} & {\footnotesize{}9.37}\tabularnewline
\hline 
\end{tabular}
\par\end{centering}

\protect\caption{Compression ratios on uniform distributed random integers.\label{tab:integer-compression}}
\end{table*}
\begin{table*}
\begin{centering}
\begin{tabular}{lccccccc}
\hline 
\multicolumn{2}{l}{{\footnotesize{}File details}} & \multicolumn{2}{c}{{\footnotesize{}FITS RICE}} & \multicolumn{2}{c}{{\footnotesize{}HDF5 LZF\_SS}} & \multicolumn{2}{c}{{\footnotesize{}HDF5 Bitshuffle}}\tabularnewline
{\footnotesize{}file name} & {\footnotesize{}size (MB)} & {\footnotesize{}ratio } & {\footnotesize{}time (s)} & {\footnotesize{}ratio} & {\footnotesize{}time (s)} & {\footnotesize{}ratio} & {\footnotesize{}time (s)}\tabularnewline
\hline 
\hline 
{\footnotesize{}photoMatchPlate-dr12.fits} & {\footnotesize{}628.3} & {\footnotesize{}2.22} & {\footnotesize{}29.30} & \textbf{\footnotesize{}2.56} & {\footnotesize{}4.76} & {\footnotesize{}1.51} & \textbf{\footnotesize{}3.24}\tabularnewline
{\footnotesize{}seguetsObjSetAllDup-0338-3138-0101.fits} & {\footnotesize{}144.4} & {\footnotesize{}1.86} & {\footnotesize{}8.49} & \textbf{\footnotesize{}4.60} & {\footnotesize{}1.99} & {\footnotesize{}1.69} & \textbf{\footnotesize{}0.96}\tabularnewline
{\footnotesize{}ssppOut-dr12.fits } & {\footnotesize{}1913.3} & {\footnotesize{}2.94} & {\footnotesize{}79.86} & \textbf{\footnotesize{}3.03} & {\footnotesize{}22.59} & {\footnotesize{}1.62} & \textbf{\footnotesize{}14.88}\tabularnewline
\hline 
\end{tabular}
\par\end{centering}

\protect\caption{Compression ratios on randomly selected SDSS DR12 binary tables data.\label{tab:SDSS-compression}}
\end{table*}
Both FITS and HDF5 support compression; FITS via the \texttt{FPACK}
utility and HDF5 via its filter pipeline. A notable feature of HDF5
is that data can be compressed and decompressed automatically and
transparently to the end user. Different filters, such as pre-compression
shuffling of data, can be chained together to enhance achieved compression
ratios. Also notable is the speed: the lossless \texttt{bitshuffle}
compression algorithm has been shown to achieve throughputs in excess
of 1GB/s on a single core while maintaining good compression ratios,
by exploiting AVX2 and SSE2 instructions that are available on x86
processors. A near-lossless version of the \texttt{bitshuffle} algorithm
has been designed specifically for radio interferometric datasets.
The CHIME pathfinder \citep{masui2015} has implemented this algorithm
for real-time data compression and storage, and boasts a compression
ratio of 3.57x, a write speed of 773 MiB/s and a read speed of 1147MiB/s.

We used the \texttt{fits2hdf} software package to create equivalent
datasets in FITS and HDF5 formats, and then compared write speed and
lossless compression ratios. On generated datasets containing random
integers, \texttt{bitshuffle} consistently outperformed FITS and standard
\texttt{GZIP} (Table \ref{tab:integer-compression}), in both compression
ratio and speed. We used random integers with bounds $(-2^{N},2^{N})$,
where $N$=7,15,23 and 31, and stored these data as 32-bit integers.
In theory, while uniformly distributed integers are non-compressible,
as the entire dynamic range of the 32-bits is not exercised it is
possible to compress these data by precisely 4x, 2x, 1.25x and 1x. 

FITS compression was performed using \texttt{FPACK}. Here, we report
the results using the default Rice compression filter with standard
options. The HDF-based filter \texttt{LZF}, used with byte-level shuffling
and scale-offset options (\texttt{LZF\_SS} in table), performed slower
than \texttt{bitshuffle} but faster than the FITS \texttt{Rice}-based
compression algorithm.

Compression tests were also performed on tabular data (Table \ref{tab:SDSS-compression}).
We selected three large (>500 MB) FITS files containing binary tables
from the Sloane Digital Sky Survey Data Release 12 (SDSS DR-12, \citealp{Alam:2015vp}),
and compared the compression of \texttt{LZF\_SS}, \texttt{bitshuffle}
and FITS \texttt{Rice} compression%
\footnote{Compression of binary tables is an experimental feature within FPACK,
enabled by the \texttt{-table} flag%
}. In our tests, \texttt{LZF\_SS} performed significantly faster than
FITS \texttt{Rice} compression and achieved a higher compression ratio,
while \texttt{bitshuffle} ran the fastest but only achieved a modest
compression ratio.

\subsection{Data access\label{sub:Data-access}}

Read speed of a file format is a major issue, which is becoming progressively
more important as average dataset sizes increase. In HDF5 is that
data can be stored in either contiguous blocks, or in discontiguous
'chunks'. if data access patterns are known, a significant speedup
in read performance can be achieved as only chunks that contain relevant
data need to be accessed; however, the entire chunk must be read.
If compression is also used, only data within the chunk being accessed
need be decompressed, and access to the raw data remains transparent
to the end user.

To benchmark real-world performance of HDF5 against FITS, we generated
a 3-dimensional dataset with dimension sizes (10000, 200, 200), consisting
of random integer32 data in the range (-$2^{23}$, $2^{23}$). We
stored these data in FITS, HDF5, and HDF5 compressed with \texttt{bitshuffle},
resulting files of size $\sim$1.5GB, $\sim$1.5GB, and $\sim$1.2GB.
For the HDF5 file, we specified a chunk size of (1000, 10, 10). During
read tests, each read was repeated 16 times and multiple copies of
the file were used so that data were not read from cache.

When reading these data back in their entirety, average read times
were 14.3, 15.7 and 13.9s, for FITS, uncompressed HDF5, and \texttt{bitshuffle}
compressed HDF5, respectively. Without chunking, HDF5 performs comparably
to FITS, indicating disk read speed as the major limitation. The read
time for a pair of randomly chosen slices along the slowest-varying
dimension (i.e. along the z-axis in a 3D data cube) were 21.1s, 0.24,
and 0.21s.

\section{Discussion\label{sec:Discussion}}

\subsection{The importance of abstraction}
%
By abstracting the data model away from file serialization, one may focus on improving
and extending the data model, without being bogged down by implementation issues. 
Our hope is that the work presented here will facilitate the creation of a robust,
community endorsed data model that is broadly applicable within astronomy. 
This data model must obey agreed standards regarding issues such as coordinate systems, 
units, and uncertainties. Another benefit of abstraction is that when next-generation 
file formats inevitably appear, the community can adapt to use them effectively, without
needing to completely rewrite existing software packages. Indeed, community discussion 
as to the future of astronomical data formats has already begun 
in earnest, see for example \citet{mink2015}. 

\subsection{HDF5 performance}

As shown in Section \ref{sub:Data-access}, HDF5 outperformed FITS
in data read speed by almost two orders of magnitude, for the case
where data along the slowest-varying axis of a data cube only is read.
In contrast, when reading the entire dataset, FITS reads slightly
faster (10\%) than chunked, uncompressed HDF5 files. While full characterization
of the effect of data access pattern upon read performance is beyond
the scope of this paper, the results presented here are representative
of best and worst-case scenarios. For any application where portions
of a larger dataset are read, a chunked HDF5 is likely to give better
read performance.

In terms of compression, the HDF5 \texttt{LZF} and \texttt{bitshuffle}
compression algorithms achieve higher throughput and compression ratios
than FITS \texttt{Rice} compression. However, we have not compared
lossy compression algorithms. When compressing floating point data,
FITS applies a pre-compression scaling filter based upon the noise
present in the image, and also applies a 'subtractive dithering' technique
\citep{2007ASPC..376..483S,2009PASP..121..414P}. The HDF5 \texttt{scaleoffset}
pre-compression filter provides similar functionality to the FITS scaling filter,
but there is currently no equivalent to the subtractive dithering.
Such functionality could be added to HDF5 by porting the subtractive dithering
algorithm from {\tt FPACK} into a HDF5 filter. Alternative compression schemes,
such as those that underlie JPEG2000, could also be ported to HDF5.

\subsection{Alternative approaches}

An important conclusion from the work presented here is that it is possible
to decouple the FITS data model from the FITS file format.
In this section, we discuss some alternative approaches and recent work on
data models within astronomy.

\paragraph{Starlink}
The Starlink package (\href{http://ascl.net/1110.012}{ascl:1110.012})
provides a utility \texttt{fits2ndf}, which converts a FITS file to
NDF format. As \citet{Jenness:2015bi} reimplemented the underlying
HDS format of NDF within HDF5, one can use \texttt{fits2ndf }to convert
a file into the new HDF5-based NDF format. NDF files are distinguished
by the attribute {\tt CLASS=NDF} in the root group. The NDF data model
defines optional array components, such as variance estimates ({\tt VARIANCE})
and pixel quality ({\tt QUALITY}), making it more extensive than the 
HDU-based data model employed in FITS. When converting a
FITS file to NDF, the {\tt fits2ndf} program stores the header cards 
as-is (i.e. as a string); this differs to the approach of HDFITS, where cards are
parsed and converted to HDF5 attributes. 

As discussed in \citet{Jenness:2014wy},
there is much to be learnt from the NDF data model, just as there is from
FITS. With both NDF and FITS data models ported to use the HDF5 file format, 
both files can be read via the HDF5 API. One then only has to consider the
differences with the data models, without concern to the minutae of the
serialization. 

\paragraph{MeasurementSets}
The MeasurementSet (MS), used by the CASA reduction package (\href{http://ascl.net/1107.013}{ascl:1107.013}), is a common file format for visibility data in radio astronomy \citep{msdef}. The MS storage model is a directory consisting of several data files nested inside child directories.
MS has no in-built compression capability, but does support chunking, caching and has a query language
for data selection. The MS standard defines data schema for images, visibility data, and single-dish
data.

\paragraph{VOTable}
The International Virtual Observatory Alliance (IVOA) VOTable%
\footnote{\url{http://www.ivoa.net/documents/VOTable/}%
} format also addresses some of the issues of FITS. VOTable is based upon
eXtensible Markup Language (XML). By itself, VOTable does not support features
such as chunking, and compression of binary data, but it can be used to store
the metadata required to setup a socket-based data stream.

\paragraph{ASDF}
Another alternative format that is currently in development is the
Advanced Science Data Format%
\footnote{\url{http://asdf-standard.readthedocs.org/en/latest/}%
} (ASDF), which combines human-readable metadata with raw binary data
(similar to FITS), and is designed primarily as an interchange format.
By design, the ASDF file format is simplistic, so lacks features available in
the more complex and abstract HDF5, such as chunking and parallel I/O support. 

\paragraph{JPEG2000}
\citet{Kitaeff:2014ua} present impressive lossless compression ratios
(3-4x) on astronomy image datasets using JPEG2000. The case for JPEG2000
is particularly compelling for cloud-based data reduction approach,
where small portions of extremely large datasets may be sent to a
client's browser via the JPEG2000 interactive protocol (JPIP). Nevertheless,
JPEG2000 is not flexible enough to store astronomy data products such
as interferometer visibilities or large data tables.

\subsection{Adding HDF5 support to existing packages}

HDF5 implements a high-level API with C, C++, Fortran, and Java interfaces.
In addition Python, IDL, Mathematica and MATLAB are all already able
to read and write HDF5 files; by extension, these languages can also
read and interpret HDFITS files. However, it should be noted that
the HDF5 API provides only a low-level approach to loading data. 

A major advantage of maintaining a FITS-like data model is that minimal
changes are required to add support for HDF5 to existing software
packages. This is as the application programming interface (API),
does not necessarily need to change. This is well evidenced by the
work of \citet{Jenness:2015bi}, who reimplemented the Hierarchical
Data System (HDS) using HDF5, by producing an API for HDF5 I/O that
is near-identical to the existing HDS API. 

\begin{figure}
\begin{centering}
\includegraphics[width=0.99\columnwidth]{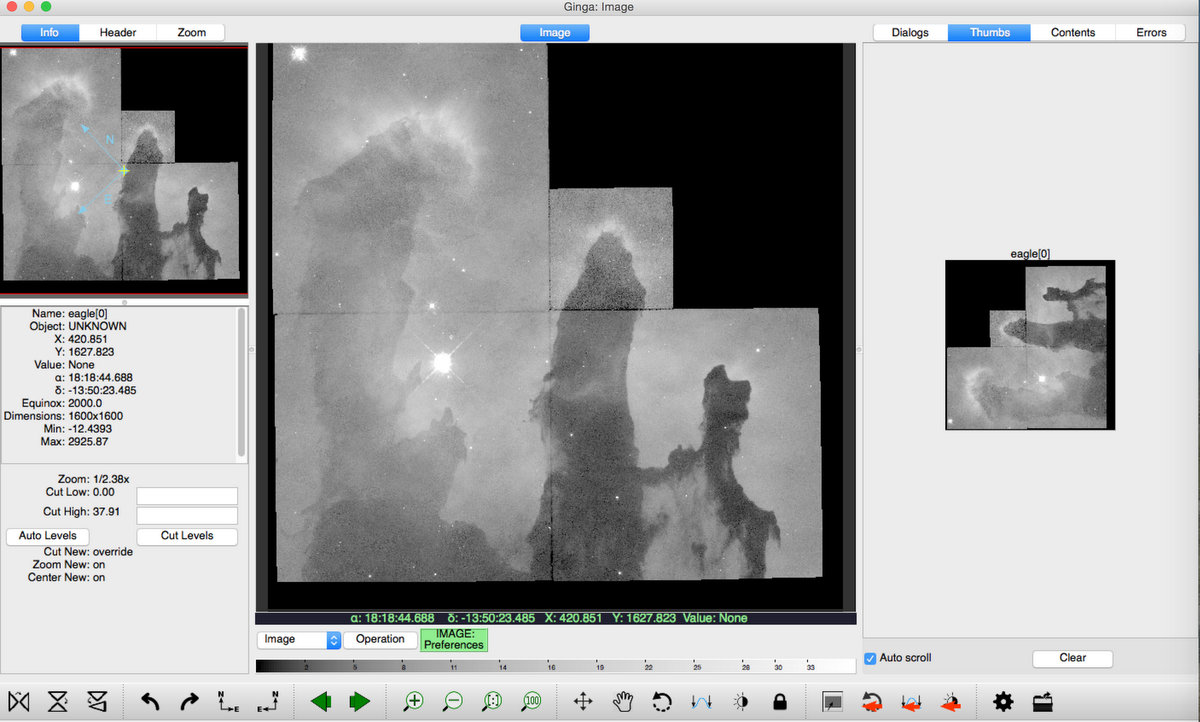}
\par\end{centering}

\protect\caption{Screenshot of HDFITS formatted data as displayed in the Ginga viewer.
The image shown is the iconic ``pillars of creation'' taken with
the Hubble Space Telescope. \label{fig:Screenshot-hdfits}}
\end{figure}

The equivalent of this for FITS would be to reimplement \texttt{CFITSIO}
to interface with HDF5 instead of FITS. We have implemented a proof-of-concept
API in \texttt{fits2hdf}, that reimplements the \texttt{open()} function
from \texttt{astropy.io.fits}. By doing so, we were able to get the
Python-based FITS viewers Ginga (\href{http://ascl.net/1303.020}{ascl:1303.020})
and Glue (\href{http://ascl.net/1402.002}{ascl:1402.002}) to read
HDFITS files simply by changing the import statement from
\begin{quote}
\texttt{\footnotesize{}from astropy.io import fits}{\footnotesize \par}
\end{quote}
to
\begin{quote}
\texttt{\footnotesize{}from fits2hdf import pyhdfits as fits}{\footnotesize \par}
\end{quote}
A screenshot of Ginga displaying HDF5 data is shown in Figure \ref{fig:Screenshot-hdfits}.

\subsection{Real-time data storage in radio astronomy}

The \texttt{fits2hdf} package is being used in the LEDA experiment
\citep{Greenhill:2012um}, to convert raw interferometric data to
\texttt{bitshuffle} compressed HDF5 files. The LEDA correlator computes
the cross product of 512 inputs of 2400 frequency channels, resulting
in an output data rate of 2.5GB per accumulation; at the current integration
length of 9s, this results in $\sim$24TB / day. Compression and file
serialization is performed in approximately 1/3 of real time, with
an achieved lossless compression ratio of $\sim$1.8 (55\% of original
size).

\subsection{Improving HDFITS}

Our motivation for HDFITS was to provide a path toward adoption of
HDF5 within the astronomy community. Extending the data model will
require input and discussion within the community, to ensure that
all needs and requirements are met.

That being said, there are several modifications that we propose would
enhance HDFITS. The first is to provide better support for columns
of data with masked values. This could perhaps be added using HDF5
dimension scales, which would also be a more flexible and appropriate
way of associating coordinate scales with a dataset. Secondly, a specification
for addressing uncertainties is sorely missing. We also suggest that
documentation (hypertext or latex), and relevant source code could
be included to enhance data provenance. With community coordination,
many of the limitations documented in \citet{Thomas:2014vq} could
be addressed.

\section{Conclusions}

The FITS file format has been an integral part of astronomy data analysis
for over 35 years. It has created an ecosystem of software that has
greatly benefitted the astronomical community. That being said, FITS
is ill-equipped to deal with the challenges that ever-increasing data
volumes impose. The proposed HDFITS standard as introduced here offers
immediate advantages, is more future-proof, and maintains the core
aspects of the FITS data model. We have shown that an HDF5-based format
achieves higher throughput and better lossless compression ratios
than FITS, and also offers faster read access via dataset chunking.

\section*{Acknowledgements}

The authors thank T. Jenness for his comments, and acknowledge support
from NSF grants AST-1106059, and OIA-1120587. 

\bibliographystyle{elsarticle-harv}
\phantomsection\addcontentsline{toc}{section}{\refname}\bibliography{hdfidi}

\newcommand{\noop}[1]{}
\begin{thebibliography}{46}
\expandafter\ifx\csname natexlab\endcsname\relax\def\natexlab#1{#1}\fi
\providecommand{\url}[1]{\texttt{#1}}
\providecommand{\href}[2]{#2}
\providecommand{\path}[1]{#1}
\providecommand{\DOIprefix}{doi:}
\providecommand{\ArXivprefix}{arXiv:}
\providecommand{\URLprefix}{URL: }
\providecommand{\Pubmedprefix}{pmid:}
\providecommand{\doi}[1]{\href{http://dx.doi.org/#1}{\path{#1}}}
\providecommand{\Pubmed}[1]{\href{pmid:#1}{\path{#1}}}
\providecommand{\bibinfo}[2]{#2}
\ifx\xfnm\relax \def\xfnm[#1]{\unskip,\space#1}\fi
\bibitem[{Abadi et~al.(2008)Abadi, Madden and Hachem}]{Abadi:2008gb}
\bibinfo{author}{Abadi, D.J.}, \bibinfo{author}{Madden, S.R.},
  \bibinfo{author}{Hachem, N.}, \bibinfo{year}{2008}.
\newblock \bibinfo{title}{{Column-stores vs. row-stores}}, in:
  \bibinfo{booktitle}{the 2008 ACM SIGMOD international conference},
  \bibinfo{publisher}{ACM Press}, \bibinfo{address}{New York, New York, USA}.
  pp. \bibinfo{pages}{967--980}.
\bibitem[{Alam et~al.(2015)Alam, Albareti, Prieto et~al.}]{Alam:2015vp}
\bibinfo{author}{Alam, S.}, \bibinfo{author}{Albareti, F.D.},
  \bibinfo{author}{Prieto, C.A.}, et~al., \bibinfo{year}{2015}.
\newblock \bibinfo{title}{{The Eleventh and Twelfth Data Releases of the Sloan
  Digital Sky Survey: Final Data from SDSS-III}}.
\newblock \bibinfo{journal}{arXiv.org}
  \href{http://arxiv.org/abs/1501.00963v2}{{\tt arXiv:1501.00963v2}}.
\bibitem[{Anderson et~al.(2010)}]{Anderson:2010tp}
\bibinfo{author}{Anderson, K.}, et~al., \bibinfo{year}{2010}.
\newblock \bibinfo{title}{{LOFAR and HDF5: Toward a New Radio Data Standard}},
  in: \bibinfo{editor}{Evans, I.}, \bibinfo{editor}{Accomazzi, A.},
  \bibinfo{editor}{Mink, D.}, \bibinfo{editor}{Rots, A.} (Eds.),
  \bibinfo{booktitle}{ADASS XX, ASP Conf. Ser. 442}, p.~\bibinfo{pages}{53}.
\bibitem[{{Astropy~Collaboration}(2015)}]{astropytables2015}
\bibinfo{author}{{Astropy~Collaboration}}, \bibinfo{year}{2015}.
\newblock
  \bibinfo{title}{\url{http://docs.astropy.org/en/v1.0/table/index.html}}.
\bibitem[{{Astropy Collaboration} et~al.(2013){Astropy Collaboration},
  {Robitaille, Thomas P.}, {Tollerud, Erik J.}, {Greenfield, Perry}
  et~al.}]{AstropyAA}
\bibinfo{author}{{Astropy Collaboration}}, \bibinfo{author}{{Robitaille, Thomas
  P.}}, \bibinfo{author}{{Tollerud, Erik J.}}, \bibinfo{author}{{Greenfield,
  Perry}}, et~al., \bibinfo{year}{2013}.
\newblock \bibinfo{title}{{Astropy: A community Python package for astronomy}}.
\newblock \bibinfo{journal}{A\&A} \bibinfo{volume}{558}, \bibinfo{pages}{A33}.
\bibitem[{Bandura et~al.(2014)}]{Bandura2014}
\bibinfo{author}{Bandura, K.}, et~al., \bibinfo{year}{2014}.
\newblock \bibinfo{title}{{Canadian Hydrogen Intensity Mapping Experiment
  (CHIME) pathfinder}}, in: \bibinfo{booktitle}{Proc. SPIE, Ground-based and
  Airborne Telescopes V}, \bibinfo{organization}{McGill Univ. (Canada)}. p.
  \bibinfo{pages}{914522}.
\bibitem[{Calabretta and Greisen(2002)}]{Calabretta:2002gz}
\bibinfo{author}{Calabretta, M.R.}, \bibinfo{author}{Greisen, E.W.},
  \bibinfo{year}{2002}.
\newblock \bibinfo{title}{{Representations of celestial coordinates in FITS}}.
\newblock \bibinfo{journal}{A\&A} \bibinfo{volume}{395},
  \bibinfo{pages}{1077--1122}.
\bibitem[{Calabretta and Roukema(2007)}]{Calabretta:2004wr}
\bibinfo{author}{Calabretta, M.R.}, \bibinfo{author}{Roukema, B.F.},
  \bibinfo{year}{2007}.
\newblock \bibinfo{title}{{Mapping on the HEALPix grid}}.
\newblock \bibinfo{journal}{MNRAS} \bibinfo{volume}{381},
  \bibinfo{pages}{865--872}.
\bibitem[{Cotton et~al.(1995)Cotton, Tody and Pence}]{Cotton1995}
\bibinfo{author}{Cotton, W.D.}, \bibinfo{author}{Tody, D.},
  \bibinfo{author}{Pence, W.D.}, \bibinfo{year}{1995}.
\newblock \bibinfo{title}{{Binary table extension to FITS.}}
\newblock \bibinfo{journal}{A\&A Supplement} \bibinfo{volume}{113},
  \bibinfo{pages}{159}.
\bibitem[{Garwood(2000)}]{garwood2000}
\bibinfo{author}{Garwood, R.W.}, \bibinfo{year}{2000}.
\newblock \bibinfo{title}{{SDFITS: A Standard for Storage and Interchange of
  Single Dish Data}}, in: \bibinfo{editor}{Bohlender, D.},
  \bibinfo{editor}{Durand, D.}, \bibinfo{editor}{Handley, T.H.} (Eds.),
  \bibinfo{booktitle}{ADASS IX, ASP Conf. Ser. 216}, p. \bibinfo{pages}{243}.
\bibitem[{Greenhill and Bernardi(2012)}]{Greenhill:2012um}
\bibinfo{author}{Greenhill, L.J.}, \bibinfo{author}{Bernardi, G.},
  \bibinfo{year}{2012}.
\newblock \bibinfo{title}{{HI Epoch of Reionization Arrays}}, in:
  \bibinfo{booktitle}{11th Asian-Pacific Regional IAU Meeting 2011, NARIT
  Conference Series}.
\bibitem[{Greisen(2003)}]{2003ASSL..285...71G}
\bibinfo{author}{Greisen, E.W.}, \bibinfo{year}{2003}.
\newblock \bibinfo{title}{{FITS: A Remarkable Achievement in Information
  Exchange}}.
\newblock \bibinfo{journal}{Information Handling in Astronomy - Historical
  Vistas. Edited by Andr{\'e} Heck} \bibinfo{volume}{285}, \bibinfo{pages}{71}.
\bibitem[{Greisen(2012)}]{Greisen:2012tw}
\bibinfo{author}{Greisen, E.W.}, \bibinfo{year}{2012}.
\newblock \bibinfo{title}{{AIPS FITS File Format}}.
\newblock \bibinfo{journal}{AIPS Memo Series, 117} .
\bibitem[{Greisen and Calabretta(2002)}]{Greisen:2002id}
\bibinfo{author}{Greisen, E.W.}, \bibinfo{author}{Calabretta, M.R.},
  \bibinfo{year}{2002}.
\newblock \bibinfo{title}{{Representations of world coordinates in FITS}}.
\newblock \bibinfo{journal}{A\&A} \bibinfo{volume}{395},
  \bibinfo{pages}{1061--1075}.
\bibitem[{Greisen et~al.(2006)Greisen, Calabretta, Valdes and
  Allen}]{Greisen:2005jp}
\bibinfo{author}{Greisen, E.W.}, \bibinfo{author}{Calabretta, M.R.},
  \bibinfo{author}{Valdes, F.G.}, \bibinfo{author}{Allen, S.L.},
  \bibinfo{year}{2006}.
\newblock \bibinfo{title}{{Representations of spectral coordinates in FITS}}.
\newblock \bibinfo{journal}{A\&A} \bibinfo{volume}{446},
  \bibinfo{pages}{747--771}.
\bibitem[{Greisen and Harten(1981)}]{1981A&AS...44..371G}
\bibinfo{author}{Greisen, E.W.}, \bibinfo{author}{Harten, R.H.},
  \bibinfo{year}{1981}.
\newblock \bibinfo{title}{{An Extension of FITS for Groups of Small Arrays of
  Data}}.
\newblock \bibinfo{journal}{A\&A Supplement} \bibinfo{volume}{44},
  \bibinfo{pages}{371}.
\bibitem[{Greisen et~al.(1980)Greisen, Wells and Harten}]{1980SPIE..264..298G}
\bibinfo{author}{Greisen, E.W.}, \bibinfo{author}{Wells, D.C.},
  \bibinfo{author}{Harten, R.H.}, \bibinfo{year}{1980}.
\newblock \bibinfo{title}{{The FITS Tape Formats - Flexible Image Transport
  Systems}}.
\newblock \bibinfo{journal}{Applications of Digital Image Processing to
  Astronomy} \bibinfo{volume}{264}, \bibinfo{pages}{298}.
\bibitem[{Griesen(2008)}]{Griesen:2008tn}
\bibinfo{author}{Griesen, E.W.}, \bibinfo{year}{2008}.
\newblock \bibinfo{title}{{The FITS Interferometry Data Interchange
  Convention}}.
\newblock \bibinfo{journal}{AIPS Memo Series, 114} .
\bibitem[{Harten et~al.(1988)}]{Harten1998}
\bibinfo{author}{Harten, R.H.}, et~al., \bibinfo{year}{1988}.
\newblock \bibinfo{title}{{The FITS tables extension}}.
\newblock \bibinfo{journal}{A\&A Supplement} \bibinfo{volume}{73},
  \bibinfo{pages}{365--372}.
\bibitem[{{HDF~Group}(2015a)}]{hdfimage2015}
\bibinfo{author}{{HDF~Group}}, \bibinfo{year}{2015}a.
\newblock
  \bibinfo{title}{\url{http://www.hdfgroup.org/HDF5/doc/ADGuide/ImageSpec.html}}.
\bibitem[{{HDF~Group}(2015b)}]{hdfgroup2015}
\bibinfo{author}{{HDF~Group}}, \bibinfo{year}{2015}b.
\newblock
  \bibinfo{title}{\url{http://www.hdfgroup.org/HDF5/doc/UG/index.html}}.
\bibitem[{{HDF~Group}(2015c)}]{hdftable2015}
\bibinfo{author}{{HDF~Group}}, \bibinfo{year}{2015}c.
\newblock
  \bibinfo{title}{\url{http://www.hdfgroup.org/HDF5/hdf5_hl/doc/RM_hdf5tb_spec.html}}.
\bibitem[{{HDF~Group}(2015d)}]{meerkat2015}
\bibinfo{author}{{HDF~Group}}, \bibinfo{year}{2015}d.
\newblock \bibinfo{title}{\url{http://www.hdfgroup.org/HDF5/users5.html}}.
\newblock \bibinfo{howpublished}{online}.
\bibitem[{Jenness(\noop{2015}\this)}]{Jenness:2015bi}
\bibinfo{author}{Jenness, T.}, \bibinfo{year}{\noop{2015}\this}.
\newblock \bibinfo{title}{{Reimplementing the Hierarchical Data System using
  HDF5}}.
\newblock \bibinfo{journal}{Astronomy and Computing}
  \href{http://arxiv.org/abs/1502.04029v1}{{\tt arXiv:1502.04029v1}}.
\bibitem[{Jenness et~al.(\noop{2015}\this)}]{Jenness:2014wy}
\bibinfo{author}{Jenness, T.}, et~al., \bibinfo{year}{\noop{2015}\this}.
\newblock \bibinfo{title}{{Learning from 25 years of the extensible
  N-Dimensional Data Format}}.
\newblock \bibinfo{journal}{Astronomy and Computing}
  \href{http://arxiv.org/abs/1410.7513}{{\tt arXiv:1410.7513}}.
\bibitem[{Jennings et~al.(1995)Jennings, Pence and Folk}]{Jennings1995}
\bibinfo{author}{Jennings, D.G.}, \bibinfo{author}{Pence, W.D.},
  \bibinfo{author}{Folk, M.}, \bibinfo{year}{1995}.
\newblock \bibinfo{title}{{Convert: Bridging the Scientific Data Format
  Chasm}}, in: \bibinfo{editor}{Shaw, R.A.}, \bibinfo{editor}{Payne, H.E.},
  \bibinfo{editor}{Hayes, J.J.E.} (Eds.), \bibinfo{booktitle}{ADASS IV, ASP
  Conf. Ser. 77}, p. \bibinfo{pages}{229}.
\bibitem[{Kemball and Wieringa(2015)}]{msdef}
\bibinfo{author}{Kemball, A.}, \bibinfo{author}{Wieringa, M.},
  \bibinfo{year}{2015}.
\newblock \bibinfo{title}{Measurementset definition version 2.0}.
\newblock \bibinfo{howpublished}{online}.
\newblock \URLprefix \url{http://casa.nrao.edu/Memos/229.html}.
\bibitem[{Kitaeff et~al.(\noop{2015}\this)Kitaeff, Cannon, Wicenec and
  Taubman}]{Kitaeff:2014ua}
\bibinfo{author}{Kitaeff, V.V.}, \bibinfo{author}{Cannon, A.},
  \bibinfo{author}{Wicenec, A.}, \bibinfo{author}{Taubman, D.},
  \bibinfo{year}{\noop{2015}\this}.
\newblock \bibinfo{title}{{Astronomical Imagery: Considerations For a
  Contemporary Approach with JPEG2000}}.
\newblock \bibinfo{journal}{Astronomy and Computing}
  \href{http://arxiv.org/abs/1403.2801}{{\tt arXiv:1403.2801}}.
\bibitem[{Masui et~al.(\noop{2015}\this)Masui, Addison, Amiri
  et~al.}]{masui2015}
\bibinfo{author}{Masui, K.}, \bibinfo{author}{Addison, G.},
  \bibinfo{author}{Amiri, M.}, et~al., \bibinfo{year}{\noop{2015}\this}.
\newblock \bibinfo{title}{{A compression scheme for radio data in high
  performance computing}}.
\newblock \bibinfo{journal}{Astronomy and Computing}
  \href{http://arxiv.org/abs/1503.00638v1}{{\tt arXiv:1503.00638v1}}.
\bibitem[{{Mink} et~al.(\noop{2015}\inpr){Mink}, {Mann}, {Hanisch}, {Rots}
  et~al.}]{mink2015}
\bibinfo{author}{{Mink}, J.}, \bibinfo{author}{{Mann}, R.G.},
  \bibinfo{author}{{Hanisch}, R.}, \bibinfo{author}{{Rots}, A.}, et~al.,
  \bibinfo{year}{\noop{2015}\inpr}.
\newblock \bibinfo{title}{{The Past, Present and Future of Astronomical Data
  Formats}}, in: \bibinfo{editor}{Taylor, A.R.}, \bibinfo{editor}{Stil, J.M.}
  (Eds.), \bibinfo{booktitle}{ADASS XXIV, ASP Con. Ser.}
\bibitem[{Pence(1999)}]{1999ASPC..172..487P}
\bibinfo{author}{Pence, W.}, \bibinfo{year}{1999}.
\newblock \bibinfo{title}{{CFITSIO, v2.0: A New Full-Featured Data Interface}},
  in: \bibinfo{editor}{Mehringer, D.M.}, \bibinfo{editor}{Plante, R.L.},
  \bibinfo{editor}{Roberts, D.A.} (Eds.), \bibinfo{booktitle}{ADASS VIII, ASP
  Conf. Ser. 172}, p. \bibinfo{pages}{487}.
\bibitem[{Pence(2002)}]{2002SPIE.4847..444P}
\bibinfo{author}{Pence, W.D.}, \bibinfo{year}{2002}.
\newblock \bibinfo{title}{{New image compression capabilities in CFITSIO}}, in:
  \bibinfo{editor}{Stark} (Ed.), \bibinfo{booktitle}{Astronomical Data Analysis
  II}, pp. \bibinfo{pages}{444--447}.
\bibitem[{Pence(2010)}]{2010ascl.soft10001P}
\bibinfo{author}{Pence, W.D.}, \bibinfo{year}{2010}.
\newblock \bibinfo{title}{{CFITSIO: A FITS File Subroutine Library}}.
\newblock \bibinfo{journal}{Astrophysics Source Code Library} ,
  \bibinfo{pages}{10001}.
\bibitem[{Pence et~al.(2009)Pence, Seaman and White}]{2009PASP..121..414P}
\bibinfo{author}{Pence, W.D.}, \bibinfo{author}{Seaman, R.},
  \bibinfo{author}{White, R.L.}, \bibinfo{year}{2009}.
\newblock \bibinfo{title}{{Lossless Astronomical Image Compression and the
  Effects of Noise}}.
\newblock \bibinfo{journal}{PASP} \bibinfo{volume}{121},
  \bibinfo{pages}{414--427}.
\bibitem[{Pence et~al.(2010)}]{2010A&A...524A..42P}
\bibinfo{author}{Pence, W.D.}, et~al., \bibinfo{year}{2010}.
\newblock \bibinfo{title}{{Definition of the Flexible Image Transport System
  (FITS), version 3.0}}.
\newblock \bibinfo{journal}{A\&A} \bibinfo{volume}{524}, \bibinfo{pages}{42}.
\bibitem[{Price et~al.(\noop{2015}\inpr)Price, Barsdell and
  Greenhill}]{Price:2014ue}
\bibinfo{author}{Price, D.C.}, \bibinfo{author}{Barsdell, B.R.},
  \bibinfo{author}{Greenhill, L.J.}, \bibinfo{year}{\noop{2015}\inpr}.
\newblock \bibinfo{title}{{Is HDF5 a good format to replace UVFITS?}}, in:
  \bibinfo{editor}{Taylor, A.R.}, \bibinfo{editor}{Stil, J.M.} (Eds.),
  \bibinfo{booktitle}{ADASS XXIV, ASP Con. Ser.}
\bibitem[{Rots et~al.(2015)Rots, Bunclark, Calabretta, Allen, Manchester and
  Thompson}]{Rots:2014iw}
\bibinfo{author}{Rots, A.H.}, \bibinfo{author}{Bunclark, P.S.},
  \bibinfo{author}{Calabretta, M.R.}, \bibinfo{author}{Allen, S.L.},
  \bibinfo{author}{Manchester, R.N.}, \bibinfo{author}{Thompson, W.T.},
  \bibinfo{year}{2015}.
\newblock \bibinfo{title}{{Representations of time coordinates in FITS}}.
\newblock \bibinfo{journal}{A\&A} \bibinfo{volume}{574}, \bibinfo{pages}{A36}.
\bibitem[{Schaaf et~al.(\noop{2015}\inpr)Schaaf, Brazier, Jenness, Nikola and
  Shepherd}]{Schaaf:2014}
\bibinfo{author}{Schaaf, R.}, \bibinfo{author}{Brazier, A.},
  \bibinfo{author}{Jenness, T.}, \bibinfo{author}{Nikola, T.},
  \bibinfo{author}{Shepherd, M.}, \bibinfo{year}{\noop{2015}\inpr}.
\newblock \bibinfo{title}{{A new HDF5 based raw data model for CCAT}}, in:
  \bibinfo{editor}{Taylor, A.R.}, \bibinfo{editor}{Stil, J.M.} (Eds.),
  \bibinfo{booktitle}{ADASS XXIV}.
\bibitem[{Seaman et~al.(2010)Seaman, Pence and White}]{2010ascl.soft10002S}
\bibinfo{author}{Seaman, R.}, \bibinfo{author}{Pence, W.},
  \bibinfo{author}{White, R.}, \bibinfo{year}{2010}.
\newblock \bibinfo{title}{{fpack: FITS Image Compression Program}}.
\newblock \bibinfo{journal}{Astrophysics Source Code Library} ,
  \bibinfo{pages}{10002}.
\bibitem[{Seaman et~al.(2007)Seaman, Pence, White, Dickinson, Valdes and
  Z{\'a}rate}]{2007ASPC..376..483S}
\bibinfo{author}{Seaman, R.}, \bibinfo{author}{Pence, W.},
  \bibinfo{author}{White, R.}, \bibinfo{author}{Dickinson, M.},
  \bibinfo{author}{Valdes, F.}, \bibinfo{author}{Z{\'a}rate, N.},
  \bibinfo{year}{2007}.
\newblock \bibinfo{title}{{Astronomical Tiled Image Compression: How and Why}},
  in: \bibinfo{editor}{Shaw, R.}, \bibinfo{editor}{Hill, F.},
  \bibinfo{editor}{Bell, D.} (Eds.), \bibinfo{booktitle}{ADASS XVI, ASP Conf.
  Ser. 376}, p. \bibinfo{pages}{483}.
\bibitem[{{SKA~Organization}(2015)}]{ska2015}
\bibinfo{author}{{SKA~Organization}}, \bibinfo{year}{2015}.
\newblock \bibinfo{title}{\url{http://www.skatelescope.org}}.
\newblock \bibinfo{howpublished}{online}.
\bibitem[{{Thomas} et~al.(2001){Thomas}, {Shaya} and {Cheung}}]{thomas2001}
\bibinfo{author}{{Thomas}, B.}, \bibinfo{author}{{Shaya}, E.},
  \bibinfo{author}{{Cheung}, C.}, \bibinfo{year}{2001}.
\newblock \bibinfo{title}{{Converting FITS into XML: Methods and Advantages}},
  in: \bibinfo{editor}{{Harnden}, Jr., F.R.}, \bibinfo{editor}{{Primini},
  F.A.}, \bibinfo{editor}{{Payne}, H.E.} (Eds.), \bibinfo{booktitle}{ADASS X,
  ASP Conf. Ser. 238}, p. \bibinfo{pages}{487}.
\bibitem[{Thomas et~al.(2014)}]{2014ASPC..485..351T}
\bibinfo{author}{Thomas, B.}, et~al., \bibinfo{year}{2014}.
\newblock \bibinfo{title}{{Significant Problems in FITS Limit Its Use in Modern
  Astronomical Research}}, in: \bibinfo{editor}{Manset, N.},
  \bibinfo{editor}{Forshay, P.} (Eds.), \bibinfo{booktitle}{ADASS XXIII, ASP
  Conf. Ser. 485}, p. \bibinfo{pages}{351}.
\bibitem[{Thomas et~al.(\noop{2015}\this)}]{Thomas:2014vq}
\bibinfo{author}{Thomas, B.}, et~al., \bibinfo{year}{\noop{2015}\this}.
\newblock \bibinfo{title}{{Learning from FITS: Limitations in use in modern
  astronomical research}}.
\newblock \bibinfo{journal}{Astronomy and Computing}
  \href{http://arxiv.org/abs/1502.00996}{{\tt arXiv:1502.00996}}.
\bibitem[{Wells and Greisen(1979)}]{1979ipia.coll..445W}
\bibinfo{author}{Wells, D.C.}, \bibinfo{author}{Greisen, E.W.},
  \bibinfo{year}{1979}.
\newblock \bibinfo{title}{{FITS - a Flexible Image Transport System}}, in:
  \bibinfo{booktitle}{5th Workshop on Image Processing in Astronomy}, p.
  \bibinfo{pages}{445}.
\bibitem[{Wise et~al.(2011)}]{2011ASPC..442..663W}
\bibinfo{author}{Wise, M.}, et~al., \bibinfo{year}{2011}.
\newblock \bibinfo{title}{{Towards HDF5: Encapsulation of Large and/or Complex
  Astronomical Data}}, in: \bibinfo{editor}{Evans, I.},
  \bibinfo{editor}{Accomazzi, A.}, \bibinfo{editor}{Mink, D.},
  \bibinfo{editor}{Rots, A.} (Eds.), \bibinfo{booktitle}{ADASS XX, ASP Conf.
  Ser. 442}, p. \bibinfo{pages}{663}.

\end{thebibliography}

\end{document}